\documentclass[10pt]{iopart}
\usepackage{graphicx}
\usepackage{dcolumn}
\usepackage{hyperref}
\usepackage{amssymb} 
\usepackage{braket}
\usepackage{tabularx}
\usepackage[percent]{overpic}
\newcolumntype{C}{>{$\displaystyle} c <{$}} 
\newcommand{\un}[1]{\ensuremath{\, \mathrm{#1}}} 

\newcommand{\tn}[1]{\mathrm{#1}}
\newcommand{\iso}[1]{$^{#1}$Ne} 
\newcommand{\lra}{\leftrightarrow} 
\newcommand{\op}[1]{\hat{#1}} 
\newcommand{\vect}[3]{\left(\begin{array}{@{}{c}@{}}#1\\#2\\#3\end{array}\right)} 
\newcommand{\dd}{\mathrm{d}} 
\newcommand{\ddd}{\hspace{-0.5ex}\mathrm{d}} 
\newcommand{\lk}{\left(}
\newcommand{\rk}{\right)}
\newcommand{\lkk}{\hspace{-2pt}\left(}
\newcommand{\lek}{\left[}
\newcommand{\rek}{\right]}
\newcommand{\tpuls}{\tau}

\newcommand{\khsp}{\hspace*{-0.7pt}}
\newcommand{\average}[1]{\overline{#1\raisebox{3mm}{}}}

\begin{document}
\jpb

\title{Coherent dynamics in a five-level atomic system}
\author{Jan Sch\"utz$^{1,2}$ 
	,
	 Alexander Martin$^1$,
	 Sanah Laschinger$^{1,3}$ 
	 and Gerhard Birkl$^1$} 
\ead{apqpub@physik.tu-darmstadt.de}
\address{$^1$Technische Universit\"at Darmstadt, Institut f\"ur Angewandte Physik, Schlossgartenstra\ss{}e 7, 64289 Darmstadt, Germany}
\address{$^2$Present address: Fraunhofer Institute for Physical Measurement Techniques IPM, Georges-K\"ohler-Allee 301, 79110 Freiburg, Germany}
\address{$^3$Present address: Bruker AXS GmbH, \"Ostliche Rheinbr\"uckenstra\ss{}e 49, 76187 Karlsruhe, Germany}
\date{\today}

\begin{abstract}
	The coherent control of multi-partite quantum systems presents one of the central prerequisites in state-of-the-art quantum information processing. With the added benefit of inherent high-fidelity detection capability, atomic quantum systems in high-energy internal states, such as metastable noble gas atoms, promote themselves as ideal candidates for advancing quantum science in fundamental aspects and technological applications. Using laser-cooled neon atoms in the metastable $^3$P$_2$ state 
	of state $1s^2 2s^2 2p^5 3s$  (LS-coupling notation) (Racah notation: $^2P_{3/2}\,3s[3/2]_2$) with five  $m_F$-sublevels, experimental methods for the preparation of all  Zeeman sublevels $\ket{m_J} = \ket{+2}, \ket{+1}, \ket{0}, \ket{-1}, \ket{-2}$ as well as the coherent control of superposition states in the five-level system  $\ket{+2}, \ldots, \ket{-2}$, in the three-level system $\ket{+2}, \ket{+1}, \ket{0}$, and in the two-level system $\ket{+2}, \ket{+1}$ are presented. The methods are based on optimized radio frequency and laser pulse sequences. The state evolution is described with a simple, semiclassical model. The coherence properties of the prepared states are studied using Ramsey and spin echo measurements.
\end{abstract}

\noindent{\it Keywords\/}: Quantum information science, Coherent dynamics, Multi-partite quantum systems, Metastable neon atoms

\maketitle
\ioptwocol

\section{Introduction \label{sec:intro}} 
Cold and ultracold atoms prepared by laser cooling and trapping techniques~\cite{Metcalf:99} present almost ideal realizations of unperturbed quantum systems. Being effectively confined in vacuum, with vanishing interactions with the environment, the atoms can be prepared in well-defined internal and external states.
Of particular interest is the preparation and manipulation of atoms in controlled superpositions of internal quantum states. 
These superposition states are important for many different fields of physics and find applications in all aspects of quantum science and technology. 
In the context of quantum information science, enhanced capabilities arise by extending the typical qubit basis to higher dimensional Hilbert space, such as qutrits, qudits, ququints etc.~\cite{Brennen2005,Smerzi2014,Jain2020}. The advantages of this approach can be fully exploited if combined with high detection efficiency on a single-particle level, as accessible with cold metastable noble gas atoms~\cite{Vassen:12}. 
Although all species of metastable noble gas atoms~\cite{Vassen:12,Kumar2011,Kale2015} can be used for this purpose, additional motivation for work with neon atoms arises from the fact that single-state preparation and the preparation of coherent superposition states have been previously studied extensively with this atomic species~\cite{Vewinger1,Vewinger2}.

In contrast to the investigation of room-temperature samples~\cite{Kumar2011,Kale2015} or atomic beams~\cite{Vewinger1,Vewinger2}, we focus our work on laser-cooled metastable atoms in order to make use of the extended interrogation times of slow and trapped atoms, the high degree of decoupling from the environment, and the best possible control over effects of Doppler-broadening and collisional interactions.
For that reason, we have chosen to investigate the coherent dynamics of laser-cooled neon atoms in the metastable $^3$P$_2$ state of the state $1s^2 2s^2 2p^5 3s$  (LS-coupling notation) (Racah notation: $^2P_{3/2}\,3s[3/2]_2$) with five $m_J$-sublevels for our work. We intend to use these results for future work in quantum information processing in high-dimensional Hilbert space an in the investigation of collisional properties in well defined interal states and their coherent superpositions~\cite{Schuetz2013b}.

The procedure for the preparation of individual Zeeman states $\ket{J, m_J}$ and their superpositions\footnote{As usual, $J$ $(F)$ denote the electronic (total) angular momentum quantum number and $m_J$ ($m_F$) the projection to the quantization axis}
usually starts with optical pumping to an extremal state with $|m_J|=J$ or to the state $m_J=0$, the degeneracy of the $m_J$-sublevels being lifted using a magnetic bias field.
Then, radio frequency (RF) or optical excitation schemes are used to prepare the target state.
Starting with the work of F. Bloch~\cite{Bloch1946}, for Zeeman states with their energy degeneracy being lifted by an external magnetic field, a comprehensive theoretical description has been established in the context of nuclear magnetic resonance. This framework has been significantly extended and generalized to optical excitations by the work of B.W. Shore~\cite{ShoreBook}. 
Since the energy splitting of the $m_J$-sublevels of every $J$-level is equidistant for small bias fields, additional precautions need to be applied in order to address individual transitions, e.\,g. utilizing the quadratic Zeeman shift~\cite{Erhard2004} or the frequency difference of transitions between different $F$-states~\cite{Law1998a}.

For atoms without hyperfine structure, such as \iso{20} and \iso{22} that are subjects of our investigations, the preparation of individual Zeeman states $\ket{J, m_J}$ and superposition states requires different preparation schemes that do not rely on the features of hyperfine structure.
For practical use, these methods need to be both efficient and robust against small fluctuations of experimental parameters like laser intensities and frequencies.
In this paper, we present experimental applications of such methods for the preparation of all Zeeman sublevels $\ket{+2}, \ket{+1}, \ket{0}, \ket{-1}, \ket{-2}$ as well as controlled superposition states using laser-cooled neon in the metastable $^3$P$_2$ state. The presented methods can readily be used for a wide range of other atomic species.

A main tool of the presented preparation methods is the application of resonant RF pulses. The resulting five-level Rabi oscillations allow for the preparation of superposition states involving all five $m_J$-states. Particularly, starting in initial state $\ket{m_J}$, the RF driven Rabi oscillation allows for the preparation of state $\ket{-m_J}$. Since Rabi oscillations in literature are usually discussed for two-level systems, we examine the five-level oscillations in greater detail in Sec.~\ref{sec:5level}, using a descriptive, semiclassical model. In Sec.~\ref{sec:2level}, we show how the Rabi oscillations can be reduced to the two-level system $(\ket{+2}, \ket{+1})$ by applying an additional light field. In Sec.~\ref{sec:3level}, we present the preparation of superposition states in the three-level system $(\ket{+2}, \ket{+1}, \ket{0})$ using fractional stimulated Raman adiabatic passage (f-STIRAP)~\cite{Bergmann1998,Vitanov_2017,Bergmann_2019,Vitanov1998}.
Similiar methods have been investigated in several fields \cite{Bergmann_2019} such as cold atoms \cite{Du_2016}, trapped ions~\cite{Wineland_1998}, spin chains~\cite{PhysRevA.88.062309}, and cavity arrays~\cite{Meher2017}.

As a first application, we use the ability to prepare the atoms in all individual $m_J$-states and controlled mixtures for measurements of the $m_J$-dependence of Penning ionization collision rates between metastable neon atoms~\cite{CopWalster2018}, the results of which are presented in a separate paper~\cite{Schuetz2013b}. Our experiments also aim at demonstrating coherent control of Penning ionization collisions using coherent superposition states~\cite{Arango2006a}.
Therefore, the coherence properties of the prepared superposition states are characterized using Ramsey and spin echo experiments in Sec.~\ref{sec:coherence}.

\section{Setup \label{sec:setup}} 
In our experiment~\cite{Schutz2012,Spoden:05,Diss-Jan13}, neon atoms are excited to the $^3$P$_2$ state in a dc discharge. The state is metastable with a lifetime of 14.73\,s~\cite{Zinner:03} and the transition $^3$P$_2$ $\lra$ $^3$D$_3$ at 640.4\,nm is used for laser cooling. A beam of metastable atoms is optically collimated, Zeeman decelerated, and captured in a magneto-optical trap. The atoms are then spin polarized to $\ket{m_J=+2}$ using circular polarized light and are transferred to an Ioffe-Pritchard magnetic trap.

While the experimental apparatus is capable of trapping all three natural occurring isotopes \iso{20}, \iso{21}, and \iso{22}~\cite{Schutz2012,Feldker2011}, the presented measurements are all performed using the most abundant \iso{20} (nuclear spin $ I = 0$), with typically $10^8$ atoms at a mean density of $10^{10}$\,cm$^{-3}$. In the magnetic trap 1-dimensional Doppler cooling along the axial direction creates an atom ensemble characterized by an axial temperature $T_z=0.2\un{mK}$ and a mean temperature of $\overline{T}=0.3\un{mK}.$

Starting point for all presented preparation schemes are free expanding ensembles of atoms in state $\ket{+2}$ released from the magnetic trap. To maintain the spin-polarization after switching off the trap, a small magnetic bias field of typically $B_0=0.38$\,G (38\,$\mu$T) is applied which defines the axis of quantization (z-axis) in the following.

We use Stern-Gerlach experiments to measure the population of the spin states of the atoms by applying a strong magnetic field gradient along the z-axis. The resulting $m_J$-dependent acceleration leads to spatial separation of the atomic ensembles in different $\ket{m_J}$, and the relative populations $p_{m_J} = |\braket{m_J|\Psi}|^2$ of state $\ket{\Psi}$ can be analyzed using absorption images of the spatially separated clouds, as shown in Fig.~\ref{fig:5level_cones}. As a guide to the eye, boxes are drawn around the density profiles of the respective states.
For a quantitative analysis, the sum of five Gaussians is fitted to the column integrated density profiles and the atom population $p_{m_J}$ is extracted for each state.

\section{RF-induced Rabi oscillations in a five-level system \label{sec:5level}} 
\begin{figure} 
	\includegraphics[width=\linewidth]{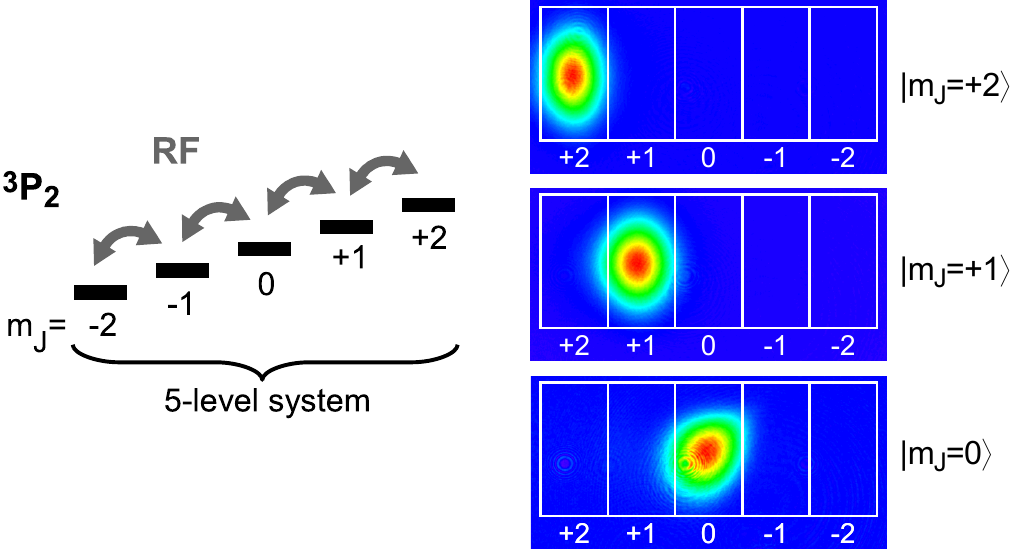}
	\caption{
		(left) Schematic of the five Zeeman sublevels $\ket{m_J}$ of the metastable state $^3$P$_2$ in a magnetic bias field and exposed to a resonant RF field.
		A small magnetic field causes an equally-spaced energy splitting of the states that are coupled by the RF field. (right) Stern-Gerlach detection of the population of the individual  $\ket{m_J}$ states by absorption imaging.
		\label{fig:skizze_5level}}
\end{figure}
The Hamiltonian of an atom with magnetic moment $\op{\mu}$ in a magnetic field $\vec{B}$ is given by
\begin{equation}\label{eq:hamilton_mu}
  \op{H}\ =\ -\op{\mu}\cdot\vec{B}\ =\ g_J\,\frac{\mu_B}{\hbar}\,\op{J}\cdot\vec{B}\,,
\end{equation}
with g-factor $g_J$, Bohr magneton $\mu_B$, reduced Planck constant $\hbar$, and angular momentum operator $\op{J}$. For a bias field $B_0$ in z-direction and a RF field $B_{\tn{RF}}\,\cos(\omega\,t)$ in x-direction, the Hamiltonian reads
\begin{eqnarray}\label{eq:hamilton}
  \op{H} 
  &=& \omega_0\,\op{J}_z + \Omega\,\cos(\omega\,t)\,\op{J}_x\,,
\end{eqnarray}
with resonance frequency $\omega_0=g_J\,\mu_B\,B_0/\hbar$ and Rabi frequency $\Omega=g_J\,\mu_B\,B_{\tn{RF}}/\hbar$. As sketched in Fig.~\ref{fig:skizze_5level}, the bias field causes an energy splitting $\Delta E=\hbar\,\omega_0\,\Delta m_J$ of the states $\ket{m_J}$ which are coupled by the RF field.

The resulting spin dynamics can easily be calculated by numerically solving the Schr\"odinger equation, but we will use a semiclassical model here for a more vivid description. In this model, which is commonly used in nuclear magnetic resonance spectroscopy~\cite{Bloch1946,Hahn1950}, magnetic moment $\vec{\mu}$ and spin $\vec{J}$ are treated as classical vectors, and the time evolution is given by the torque equation
\begin{eqnarray}
  \frac{\dd}{\dd t}\vec{J}\ &=&\ \vec{\mu}\times\vec{B}
  \ =\ g_J\,\frac{\mu_B}{\hbar}\,\vec{J}\times\vec{B}\\
  &=&\ \vect{J_x}{J_y}{J_z} \times  \vect{\Omega\cos(\omega\,t)}{0}{\omega_0}.
\end{eqnarray}
It is convenient to transform from the laboratory coordinate system (unit vectors: $\vec{e}_x$, $\vec{e}_y$, $\vec{e}_z$) to a coordinate system $(\vec{e}_{\tilde{x}}, \vec{e}_{\tilde{y}}, \vec{e}_{\tilde{z}})$ that rotates with frequency $\omega$ about the x-axis, so that the torque equation becomes
\begin{equation}\label{eq:rot_full}
  \frac{\dd}{\dd t}\vec{J}
  \ =\ \vect{ J_{\tilde{x}} }{ J_{\tilde{y}} }{ J_{\tilde{z}} }
    \times \vect{\Omega/2\,[1+\cos(2\,\omega\,t)]}{\Omega/2\,\sin(2\,\omega\,t)}{\omega_0-\omega}.
\end{equation}
In many practical applications, $\omega$ is much larger than $\Omega$ and a rotating wave approximation (RWA) can be performed, i.\,e. the terms oscillating with frequency $2\omega$ in Eq.~\eref{eq:rot_full} can be neglected, resulting in
\begin{equation}\label{eq:rot_RWA}
  \frac{\dd}{\dd t}\vec{J}
  \ \approx\ \vect{ J_{\tilde{x}} }{ J_{\tilde{y}} }{ J_{\tilde{z}} }
    \times \vect{\Omega/2}{0}{\omega_0-\omega}.
\end{equation}
Thus, for a resonant RF field $(\omega=\omega_0)$, the time evolution of $\vec{J}$ simply corresponds to a rotation about $\vec{e}_{\tilde{x}}$ by angle $\theta(t)=\Omega\,t/2$.
Of special importance are the so called $\pi$-pulse $(\theta=\pi)$ and $\pi/2$-pulse $(\theta=\pi/2)$. For $\vec{J}$ initially aligned in direction of $\vec{e}_z$, the $\pi$-pulse yields an alignment in direction of $-\vec{e}_z$ and the $\pi/2$-pulse an alignment in the x-y-plane, respectively.
The time evolution without RF field $(\Omega=0)$ corresponds to a static state in the rotating frame and to a rotation about the z-axis in the laboratory frame.

\begin{figure*}
  \includegraphics[width=\linewidth]{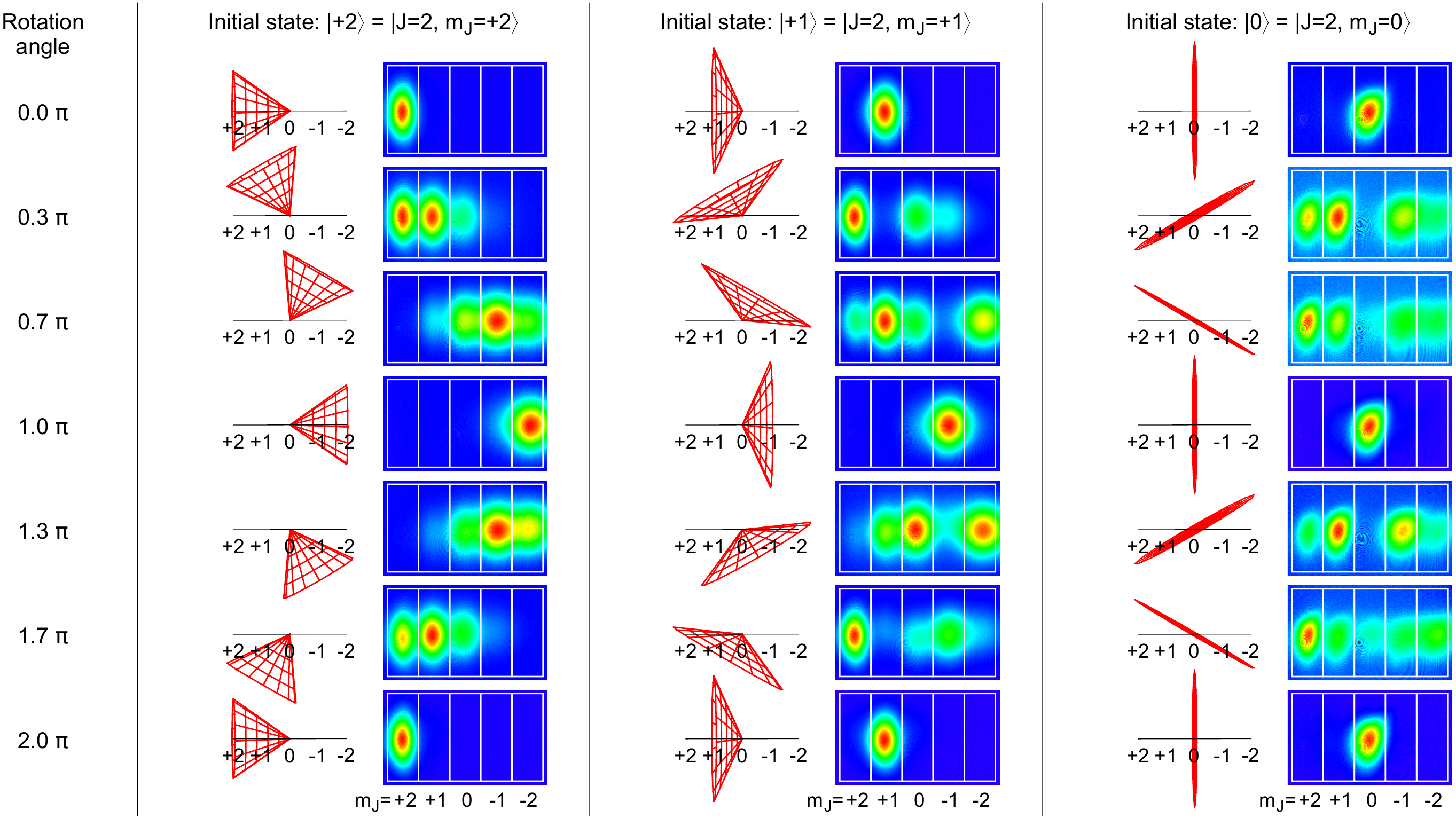}
  \caption{Absorption images of Stern-Gerlach separated ensembles and corresponding spin cones after application of resonant RF pulses for initial states $\ket{\psi_0}=\ket{+2}$, $\ket{+1}$, and $\ket{0}$. The images are arranged as a function of rotation angle $\theta(t)=\Omega\,t/2$ about the $\tilde{x}$-axis.
  	\label{fig:5level_cones}}
\end{figure*}
\begin{table*}
  \caption{Relative populations $p_{m_J}(\theta) = |\bra{m_J}\hat{D}_{\tilde{x}}(\theta)\ket{\psi_0}|^2$ for rotation of initial states $\ket{\psi_0}=\ket{+2}$, $\ket{+1}$, and $\ket{0}$ by angle $\theta$ about $\vec{e}_{\tilde{x}}$ (resonant RF field). \label{tab:states_analytic}}
    \renewcommand{\arraystretch}{1.5}
    \begin{indented}
    \item[]
    \begin{tabularx}{\textwidth}{@{\extracolsep{\fill}}CCCC}
       \br
      & \ket{\psi_0}=\ket{+2} & \ket{\psi_0}=\ket{+1} & \ket{\psi_0}=\ket{0} \\
      \mr
   p_{+2}(\theta) &
     \cos^8(\theta/2) & 4\,\cos^6(\theta/2)\,\sin^2(\theta/2) & 3/8\,\sin^4(\theta) \\
   p_{+1}(\theta) &
     4\,\cos^6(\theta/2)\,\sin^2(\theta/2) & \cos^4(\theta/2)\,[2\,\cos(\theta)-1]^2 & 3/2\,\cos^2(\theta)\,\sin^2(\theta) \\

   p_{0}(\theta) &
     3/8\,\sin^4(\theta) & 3/2\,\cos^2(\theta)\,\sin^2(\theta) & 1/16\,[1 + 3 \cos(2\theta)]^2 \\
   p_{-1}(\theta) &
     4\,\cos^2(\theta/2)\,\sin^6(\theta/2) & \sin^4(\theta/2)\,[2\,\cos(\theta)+1]^2 & 3/2\,\cos^2(\theta)\,\sin^2(\theta) \\
   p_{-2}(\theta) &
     \sin^8(\theta/2) & 4\,\cos^2(\theta/2)\,\sin^6(\theta/2) & 3/8\,\sin^4(\theta) \\
    \br
    \end{tabularx}
    \end{indented}
\end{table*}
With the Stern-Gerlach experiments we measure the projection of the rotated state to the z-axis. To calculate the projection, $\vec{J}$ needs to be treated quantum mechanically again. In a resonant RF field, following Eq.~\eref{eq:rot_RWA}, the initial state  
\begin{equation}
\ket{\Psi_0}=\sum_{m_J}c_{m_J}(0)\ket{m_J}
\end{equation}
is transformed to the rotated state
\begin{equation}
  \ket{\Psi(\theta)}=\op{D}_{\tilde{x}}(\theta)\ket{\Psi_0},
\end{equation}
with the rotation operator
\begin{equation}
  \op{D}_{\tilde{x}}(\theta) = \exp\left(-\frac{i}{\hbar}\theta\,\vec{e}_{\tilde{x}}\cdot\op{J}\right).
\end{equation}
The resulting relative populations $p_{m_J}(\theta)=|\bra{m_J}$ $\hat{D}_{\tilde{x}}(\theta)\ket{\psi_0}|^2$ for initial states $\ket{\psi_0}=\ket{+2}$, $\ket{+1}$, and $\ket{0}$ are given in Tab.~\ref{tab:states_analytic}.

For the quantum state $\ket{J, m_J}$, only the square absolute value $\op{J}^2\ket{J, m_J}= J(J+1)\hbar^2\ket{J, m_J}$ and the z-component $\op{J}_z\ket{J, m_J}=m_J\hbar\ket{J, m_J}$ are well defined, but the x- and y-component are uncertain. Therefore, instead of a vector, the quantum state $\ket{J, m_J}$ is better illustrated by a \emph{cone} with height $m_J$ and base radius $\sqrt{J(J+1)-m^2_J}$ in an abstract three-dimensional vector space $(j_x, j_y, j_z)$, and the spin dynamics corresponds to a rotation of the cone~\cite{Cook1979a}. For states with $m_j=0$, this results in rotation of a flat disk with radius $\sqrt{J(J+1)}$.

In Fig.~\ref{fig:5level_cones}, absorption images of Stern-Gerlach separated ensembles after application of resonant RF pulses with $\omega_0=2\pi\times(800\pm10)$\,kHz and $\Omega=2\pi\times(95\pm5)$\,kHz are plotted for initial states $\ket{\Psi_0}=\ket{+2}$, $\ket{+1}$, and $\ket{0}$.
(It is shown in the following sections how to prepare states $\ket{+1}$ and $\ket{0}$ in the first place).
In this case, the RWA can be applied. To visualize the dynamics, also the corresponding spin cones are sketched. 
The images are arranged as a function of rotation angle $\theta(t)=\Omega\,t/2$ about the $\tilde{x}$-axis.
Starting in $\ket{m_J}$, the rotation results in superpositions involving all five $m_J$-states, reaching maximum population in $\ket{-m_J}$ after a rotation by $\pi$, and in $\ket{m_J}$ again after a rotation by $2\pi$. In case of state $\ket{0}$ the maximum population in the initial state is already reached after a rotation by $\pi$.

\begin{figure}
  \includegraphics[width=\linewidth]{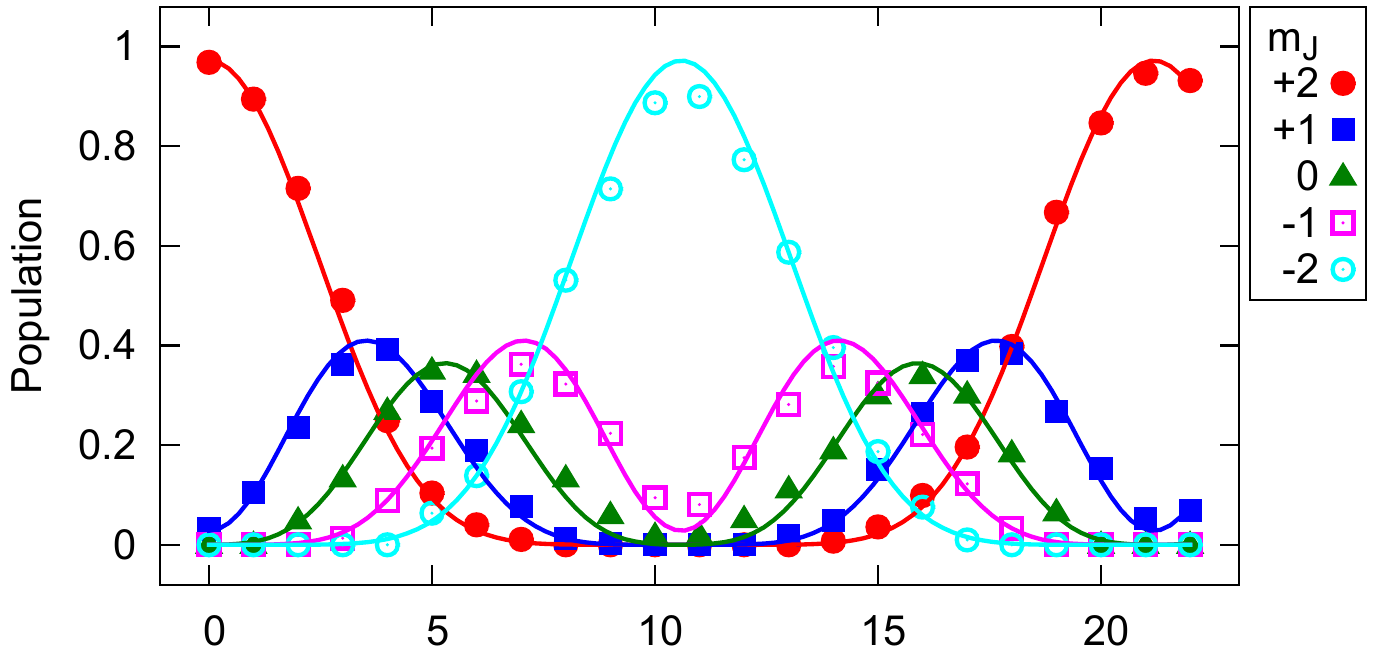}
  \includegraphics[width=\linewidth]{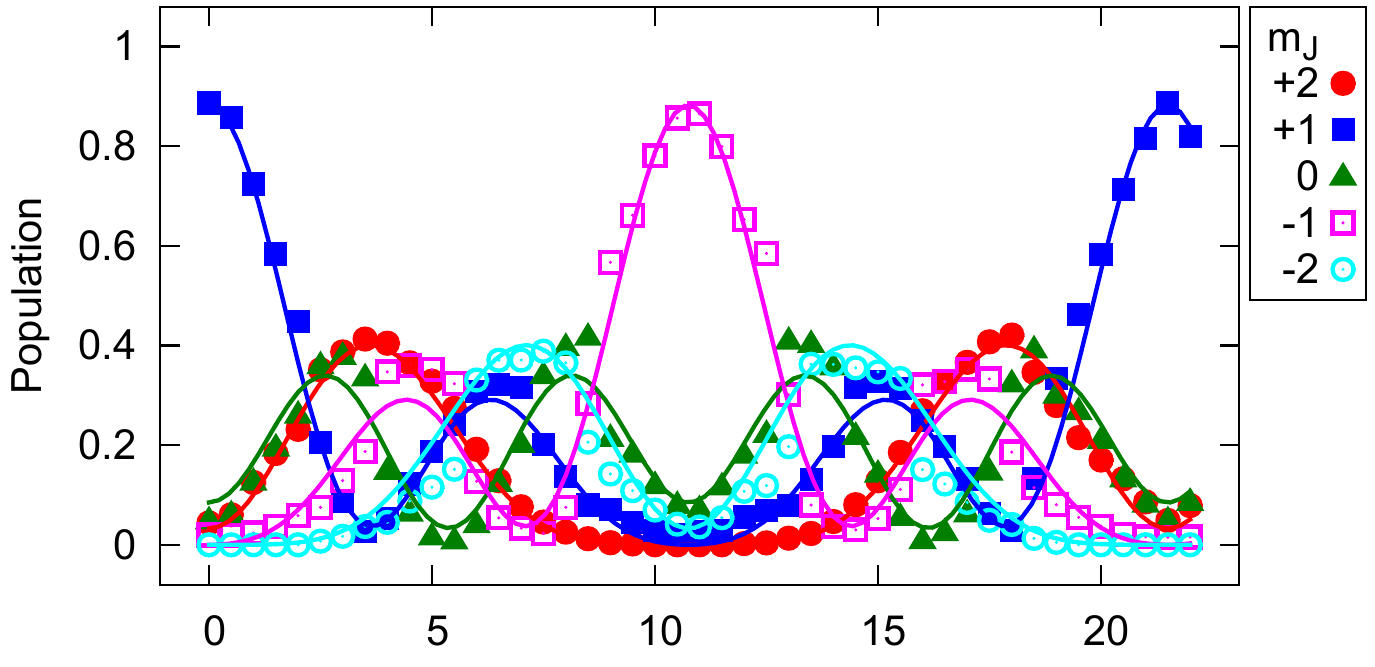}
  \includegraphics[width=\linewidth]{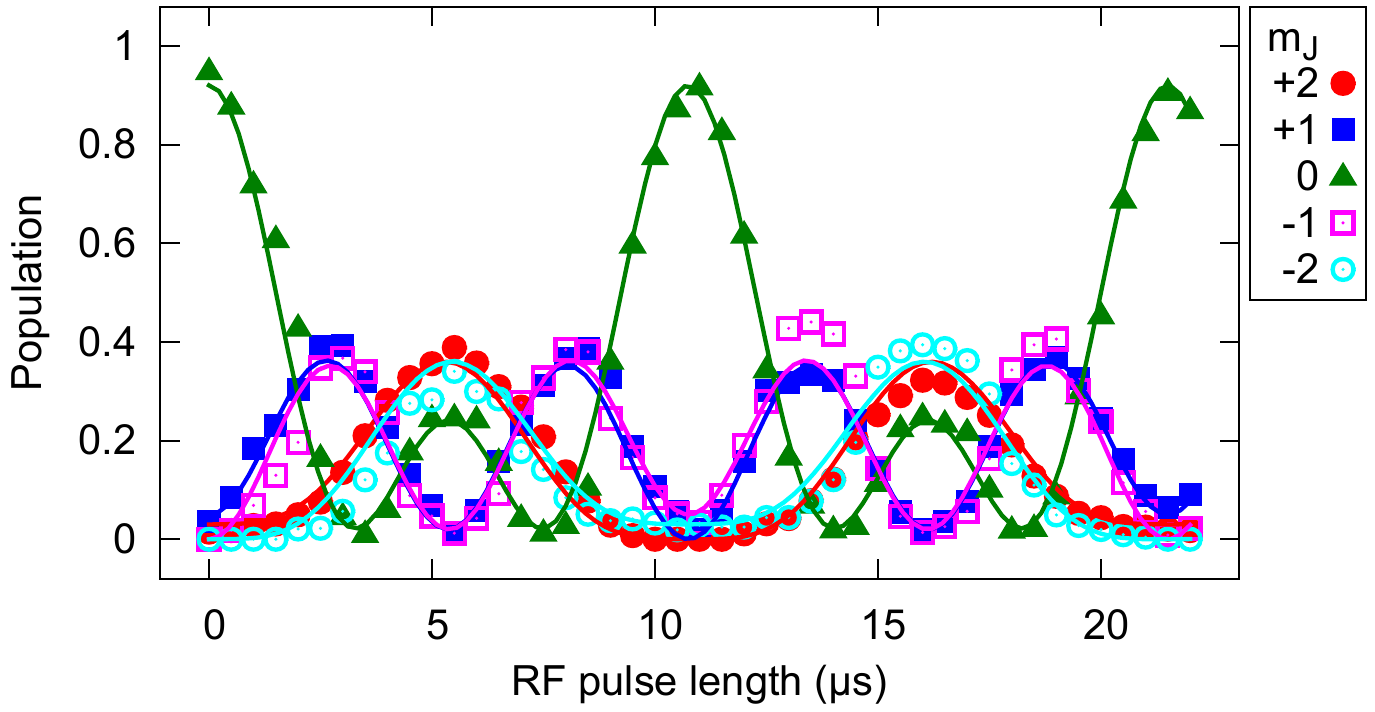}
  \caption{Relative populations $p_{m_J}$ after application of resonant RF pulses with $\omega=\omega_0=2\pi\times(800\pm10)$\,kHz and $\Omega=2\pi\times(95\pm5)$\,kHz for initial states $\ket{\psi_0}=\ket{+2}$ (top), $\ket{+1}$ (center), and $\ket{0}$ (bottom). The size of the symbols corresponds to the measurement uncertainties. The lines represent the fitted functions given by Tab.~\ref{tab:states_analytic} with applied RWA. \label{fig:5level}}
\end{figure}
The corresponding relative populations $p_{m_J}(t)$, determined by fitting Gaussians to the column integrated density profiles of the absorption images, are plotted in Fig.~\ref{fig:5level} as a function of RF pulse length $t$.
The functions given by Tab.~\ref{tab:states_analytic} have been fitted to the data points, varying $\Omega$ and the initial populations $p_{m_J}(0)$.
For the atoms released from the trap, without further manipulation, we find relative populations of $p_{+2}(0)=0.97\pm0.02$ and $p_{+1}(0)=0.03\pm0.02$. 
The preparation of $\ket{+1}$ and $\ket{0}$ leads to a small decrease in the degree of polarization, and the measured relative populations $p_{+1}(0)$ or $p_{0}(0)$ are typically between 0.90 and 0.95, respectively.

\begin{figure} 
  \includegraphics[width=\linewidth]{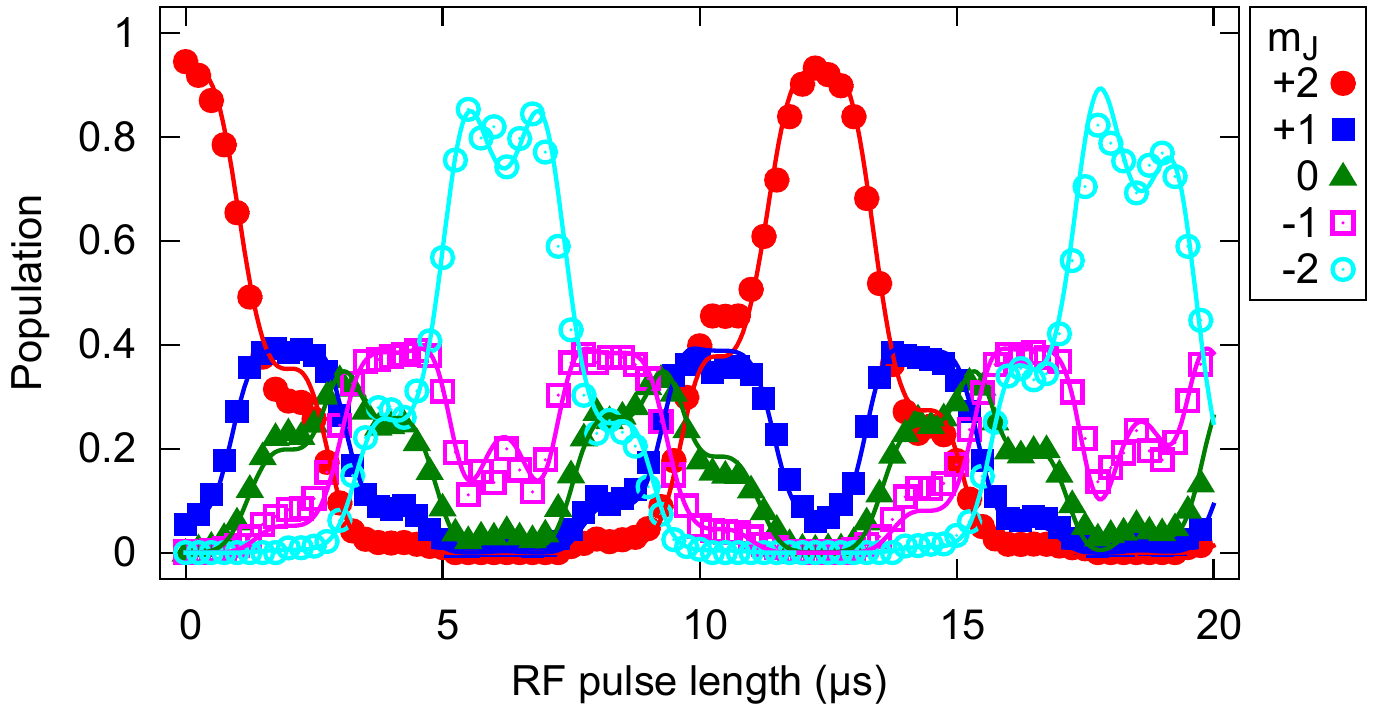}
  \caption{Relative populations $p_{m_J}$ after application of resonant RF pulses with $\omega=\omega_0=2\pi\times242$\,kHz and $\Omega=2\pi\times160$\,kHz. The lines represent a numerical solution of Eq.~\eref{eq:hamilton}, i.\,e. without RWA. \label{fig:non_rwa}}
\end{figure}

In the measurements presented in Fig.~\ref{fig:5level}, the resonance frequency $\omega_0$ $(=\omega)$ is much larger than the Rabi frequency $\Omega$. Thus, the RWA can be applied in Eq.~\eref{eq:rot_full} and the spin dynamics results in the expected five-level Rabi oscillations. 
A different situation is presented in Fig.~\ref{fig:non_rwa} where $\omega_0=2\pi\times242$\,kHz and $\Omega=2\pi\times160$\,kHz are of similar value, which is realized by lowering the bias field and increasing the RF power. As a consequence, the RWA cannot be applied in this case.
The simple Rabi oscillation is now superimposed by an oscillation of frequency $2\,\omega$. The time evolution of the relative populations is well described by a numerical solution of Eq.~\eref{eq:hamilton}, with initial populations $p_{+2}(0)=0.93$ and $p_{+1}(0)=0.07$, that is also plotted in Fig.~\ref{fig:non_rwa}.

%

\section{\texorpdfstring{Reduction to the two-level system $(\ket{+2}, \ket{+1})$}{Reduction to the two-level system (+2, +1)}\label{sec:2level}} 
Applying an additional light field, the RF driven five-level Rabi oscillation can be reduced to the two-level system $(\ket{+2}, \ket{+1})$ in a robust fashion. Particularly, this can be used to prepare state $\ket{+1}$.
The idea is to use the light induced AC Stark shift to introduce an asymmetry to the equally spaced Zeeman splitting which allows for the addressing of individual transitions $\ket{m_J}\lra\ket{m_J+1}$.

\begin{figure} 
  \includegraphics[width=\linewidth]{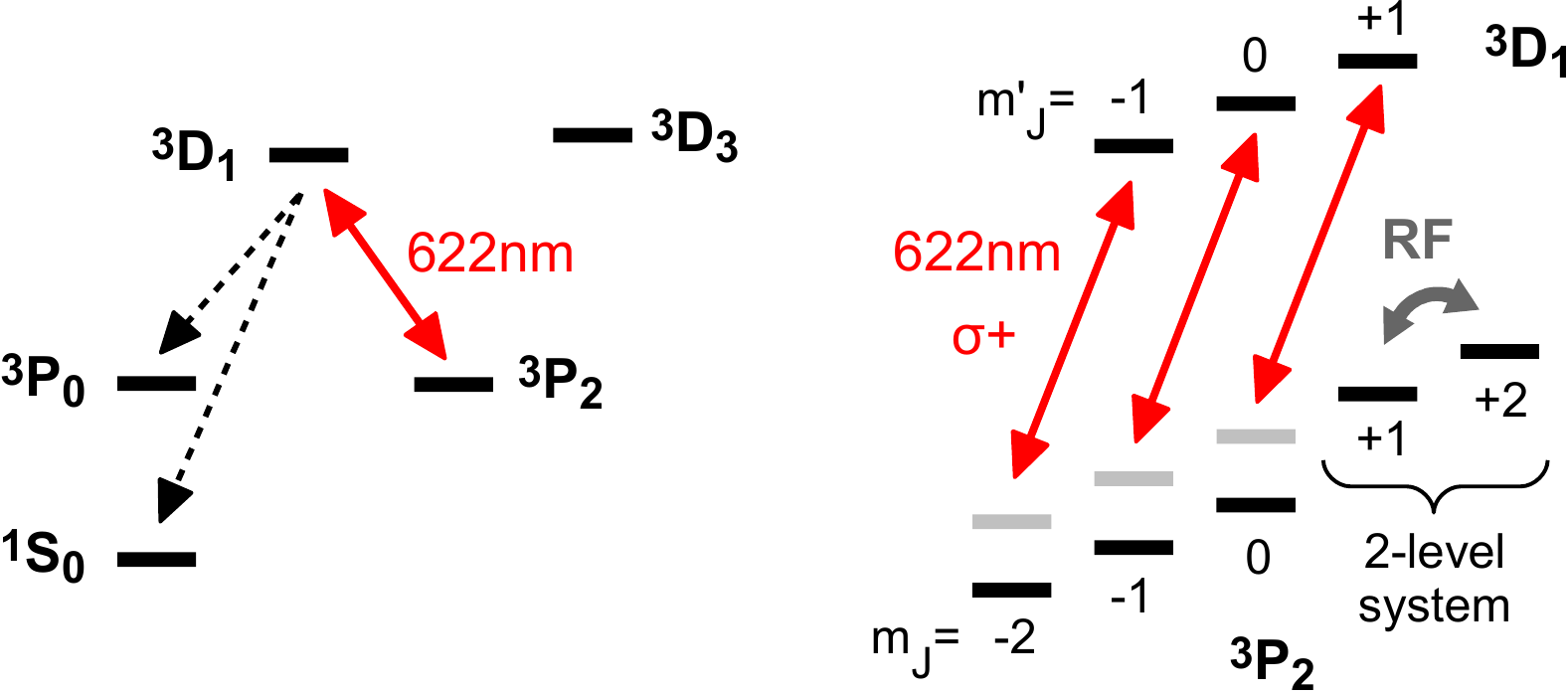}
  \caption{Reduction of the RF driven Rabi oscillations to a two-level system. (left) Light at 622\,nm couples the metastable state $^3$P$_2$ to state $^3$D$_1$ that decays with large probability to states $^3$P$_0$ and $^1$S$_0$. (right) The circularly polarized light only couples to states with $m_J<1$ and causes an AC Stark shift of the corresponding levels. Thus, RF with frequency $\omega_0$ is solely resonant to transition $\ket{+2}\lra\ket{+1}$.  \label{fig:2level_transitions}}
\end{figure}

We use laser light at 622\,nm that couples the metastable state $^3$P$_2$ to state $^3$D$_1$ which rapidly decays to other states and, thus, must not be populated (Fig.~\ref{fig:2level_transitions}, left).
The light is circularly polarized and couples only to states $\ket{0}$, $\ket{-1}$, and $\ket{-2}$, resulting in an AC Stark shift of the respective levels that shall not be populated in the following (Fig.~\ref{fig:2level_transitions}, right).
The frequency is detuned by $\Delta=-2\pi\times130$\,MHz from resonance (linewidth $\Gamma_{622}= 2\pi\times1.01$\,MHz) in order to suppress unwanted excitation of $^3$D$_1$ due to imperfect polarization of the light.

Because of the AC Stark shift, the RF field with frequency $\omega_0$ is only resonant to transition $\ket{+2}\lra\ket{+1}$, and the Rabi oscillation is reduced to the respective two-level system (Fig.~\ref{fig:2level_transitions}, right).
The exact value of the Stark shift, that might vary over the extend of the cloud, is not of importance. The shift just has to be sufficiently large to suppress the excitation of state $\ket{0}$.

\begin{figure}
	\includegraphics[width=\linewidth]{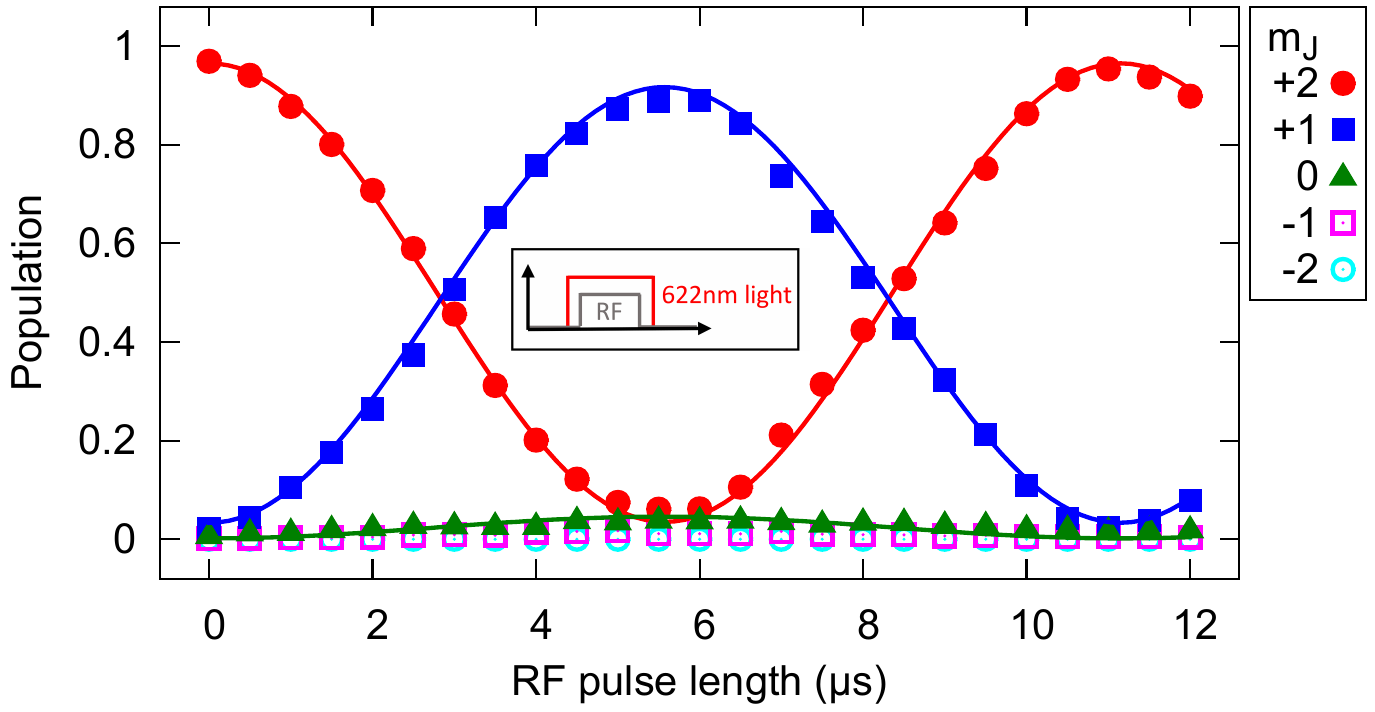}
	\caption{RF driven Rabi oscillation with simultaneous application of circularly polarized light at 622\,nm. The oscillation is reduced to the two level system $(\ket{+2}, \ket{+1})$. \label{fig:2level}}
\end{figure}

In Fig.~\ref{fig:2level}, the measured relative populations are plotted as a function of RF pulse length $t$. The oscillation is essentially reduced to the two-level system $(\ket{+2}, \ket{+1})$, with less than 5\,\% relative population in other states.
Starting in state $\ket{+2}$ ($p_{+2}=0.97,~p_{+1}=0.03$), a maximum transfer $\ket{+2}\rightarrow\ket{+1}$ is measured at $t=5.5\,\mu$s  ($p_{+2}=0.06,~p_{+1}=0.89,~p_{0}=0.05$, with uncertainties $\Delta p_{m_J}=\pm0.02$).
The relative population of state $\ket{+2}$ follows the two-level oscillation formula
\begin{equation}\label{eq:2level}
  p_{+2}(t) = p_{+2}(0)\,\cos^2\lk\frac{\Omega\,t}{2}\rk
  \ +\ p_{+1}(0)\,\,\sin^2\lk\frac{\Omega\,t}{2}\rk,
\end{equation}
with $\Omega=2\pi\times90$\,kHz, $p_{+2}(0)=0.97$, and $p_{+1}(0)=0.03$.
The populations of $\ket{+1}$ and $\ket{0}$ are well described by the assumption $p_{0}(t)=0.05\,p_{+1}(t)$, with $p_{+2}(t)+p_{+1}(t)+p_0(t)=1$. The resulting functions are also plotted in Fig.~\ref{fig:2level}. The loss of metastable atoms due to excitation of $^3$D$_1$ is found to be less than 3\,\%.

\section{\texorpdfstring{Three-level system $(\ket{+2}, \ket{+1}, \ket{0})$ using fractional STIRAP}{Three-level system (+2, +1, 0) using fractional STIRAP} \label{sec:3level}} 
In addition to pure RF coupling, STIRAP (stimulated Raman adiabatic passage) has been applied for optical coherent state transfer in many systems with great success~\cite{Bergmann1998,Vitanov_2017,Bergmann_2019}.
STIRAP exploits the dark state of a $\Lambda$-system, that is coupled by two laser pulses, for a (nearly) complete population transfer between the two unexcited states.
The \emph{pump} pulse couples the initially populated state to the excited state, and the \emph{Stokes} pulse couples the initially unpopulated state to the excited state.
Counterintuitively, the Stokes pulse precedes the pump pulse.
The main advantage of the population transfer via STIRAP, compared to other multi-photon processes, is its robustness against moderate fluctuations of experimental parameters, such as laser power and laser frequency.
The only strict demands that have to be fulfilled are the two-photon resonance and the adiabaticity criterion. 

In this section, we demonstrate how STIRAP can be applied for our purpose and show that it can extend the set of available tools and accessible states. Significant advantages of STIRAP arise from the fact that unidirectional state transfer and the relieable population of a specific state become possible. On the other hand, the unintenional population transfer to higher lying states exhibiting spontaneous decay has to be avoided. This is not the case for pure RF state transfer but can be minimized for STIRAP as well by following the established protocols. 

\begin{figure} 
  \includegraphics[width=\linewidth]{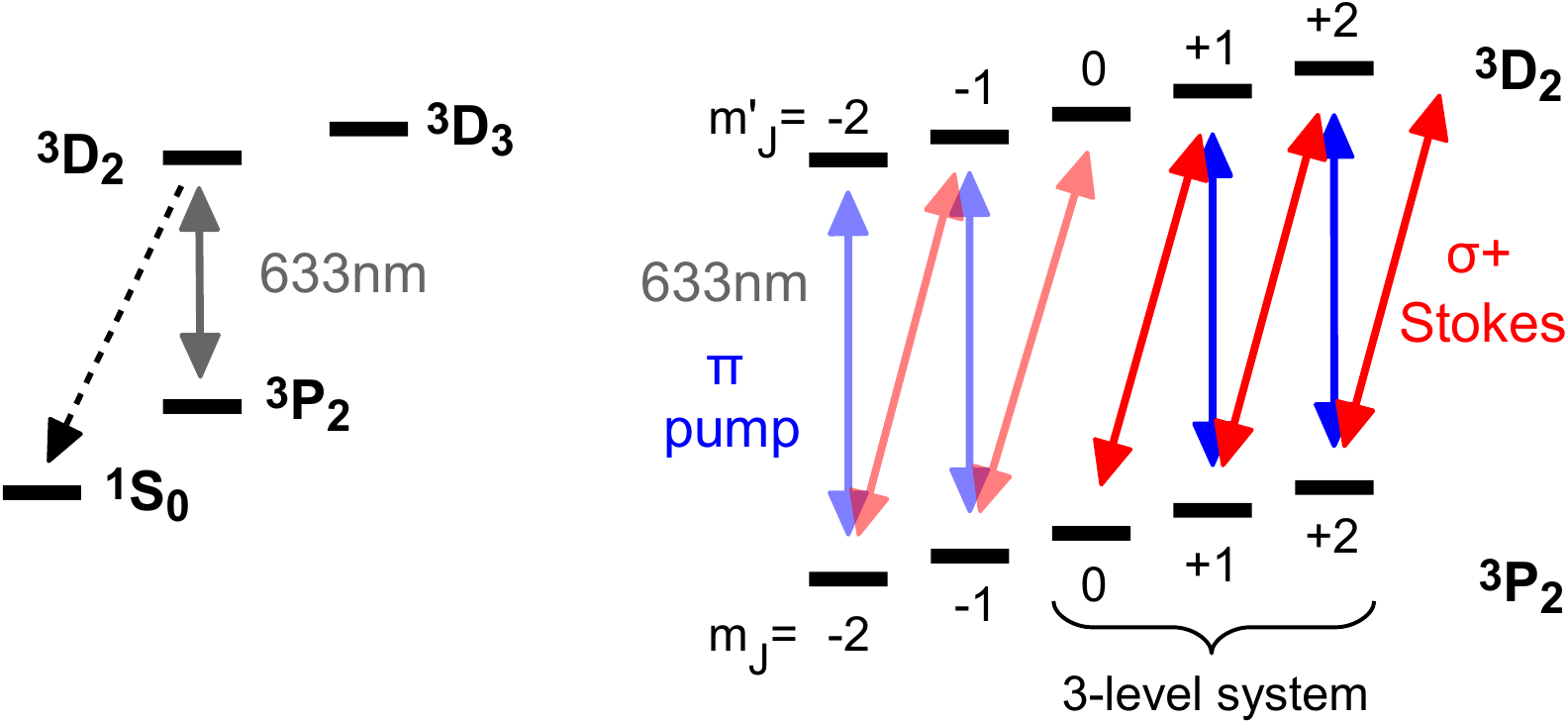}
  \caption{Coupling scheme used for STIRAP. (left) Light at 633\,nm couples state $^3$P$_2$ to state $^3$D$_2$ which decays with large probability to the ground state and, thus, must not be populated. (right) The Stokes pulse does not couple to state $\ket{+2}$, and, since the transition $\ket{J=2, m_J=0}\lra\ket{J'=2, m'_J=0}$ is dipole forbidden, the pump pulse does not couple to $\ket{0}$. Thus, the population transfer takes place in the system $(\ket{+2}, \ket{+1}, \ket{0})$. \label{fig:stirap_transitions}}
\end{figure}

In this work, we use STIRAP to prepare state $\ket{0}$ and fractional STIRAP (f-STIRAP)~\cite{Vitanov1998}, a simple extension of the usual STIRAP sequence, to prepare coherent superpositions of states $\ket{+2}$, $\ket{+1}$, and $\ket{0}$.
We use light at 633\,nm that couples state $^3$P$_2$ to state $^3$D$_2$ (Fig.~\ref{fig:stirap_transitions}, left). The light is derived from a single dye laser and split into two beams,
the polarization and direction of which are chosen to excite $\pi$-transitions (pump) and $\sigma^+$-transitions (Stokes), respectively.
As shown in Fig.~\ref{fig:stirap_transitions} (right), the Stokes light does not couple the initially populated state $\ket{+2}$ to an excited state, nor does the pump light couple state $\ket{0}$ to an excited state. 
Thus, starting in $\ket{+2}$, the linkage pattern ends in $\ket{0}$, and STIRAP results in a population transfer $\ket{+2}\rightarrow\ket{0}$, with $\ket{+1}$ being populated as an intermediate state.

Acousto-optic modulators (AOMs) are used to form pulses of variable form and length.
For STIRAP, we use Gaussian pulses corresponding to a time-dependent Stokes-pulse Rabi frequency
\begin{equation}\label{eq:STIRAP_stokes}
  \Omega_{\tn{S}}(t) = \Omega_0 \exp\lk\frac{-t^2}{\tpuls^2}\rk
\end{equation}
and time-dependent pump-pulse Rabi frequency
\begin{equation}\label{eq:STIRAP_pump}
  \Omega_{\tn{P}}(t) = \Omega_{0} \exp\lk\frac{-(t-\delta\khsp t)^2}{\tpuls^2}\rk,
\end{equation}
with pulse lengths $\tpuls=0.55\,\mu$s and maximum Rabi frequencies of typically $\Omega_0\approx2\pi\times40$\,MHz at the beam centers.
The light is detuned by $\Delta=2\pi\times20\,\tn{MHz}\approx8\,\Gamma_{622}$ from resonance to reduce spontaneous scattering of photons with deviating polarization due to imperfect polarization of pump or Stokes light which may couple the dark state to the exited state. 
The AOMs are also used to compensate the Zeeman shift $(\Delta_{\tn{S}}=\Delta_{\tn{P}}+\omega_0)$.

For a counterintuitive pulse order, with delay $\delta t$ in the range $0.4\,\mu\tn{s}\leq\delta t\leq 1\,\mu\tn{s}$, we measure a robust population transfer $\ket{+2}\rightarrow\ket{0}$ with final relative population $p_0=0.94$.
Though, 40\,\% of the metastable atoms are lost during the transfer by excitation and spontaneous decay of state $^3$D$_2$.
This can be assigned to the different Doppler shifts of the two STIRAP light fields, seen by the moving atoms, that result in a detuning from two-photon resonance. 
The atom populations after STIRAP present the initial states of the respective Rabi oscillation shown in Figs.~\ref{fig:5level_cones} and \ref{fig:5level}.

The STIRAP process can be robustly interrupted in order to prepare coherent superpositions of states $\ket{+2}$, $\ket{+1}$, and $\ket{0}$, by a method which is called fractional STIRAP or f-STIRAP~\cite{Vitanov1998}. 
Here, the Stokes pulse is extended by a second pulse that has the same time dependence as the pump pulse,
%
\begin{equation}\label{eq:fSTIRAP_pulses}
  \Omega_{\tn{S}}(t) = \Omega_0 \exp\lk\frac{-t^2}{\tpuls^2}\rk
  + \eta\,\Omega_{0} \exp\lk\frac{-(t-\delta\khsp t)^2}{\tpuls^2}\rk\,.
\end{equation}
Since pump and Stokes pulse vanish simultaneously for $t\rightarrow+\infty$, with constant ratio $\Omega_{\tn{S}}(t)/\Omega_{\tn{P}}(t)=\eta$, the dark state, that is populated during the STIRAP process, is adiabatically mapped to the atomic superposition state.
Varying $\eta$, the mixing angle of the dark state can be chosen, resulting in relative populations \cite{Vitanov1998}
\begin{eqnarray}
\label{eq:fSTIRAP_pops_a}
  p_{+2}(\eta) &=& \frac{3\,\eta^4}{2+6\eta^2+3\eta^4}\,,\\
\label{eq:fSTIRAP_pops_b}
  p_{+1}(\eta) &=& \frac{6\,\eta^2}{2+6\eta^2+3\eta^4}\,,\\
\label{eq:fSTIRAP_pops_c}
  \tn{and}~ p_{0}(\eta) &=& \frac{2}{2+6\eta^2+3\eta^4}\,.
\end{eqnarray}

\begin{figure}
	\includegraphics[width=\linewidth]{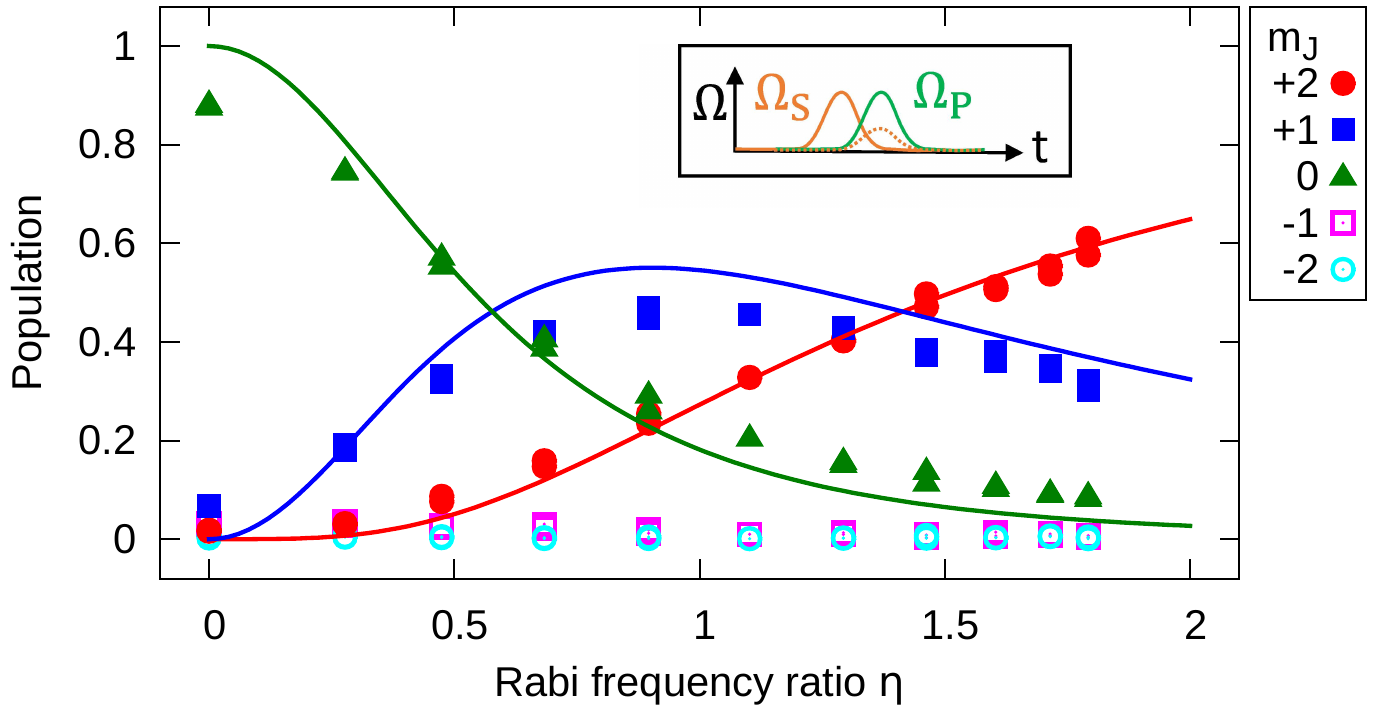}
	\caption{Relative populations after f-STIRAP as a function of the ratio $\eta$ of the Rabi frequencies with $\tau=0.55\,\mu$s, $\delta t=0.7\,\mu$s, and $\Omega_0\approx2\pi\times40\,$MHz. The lines correspond to calculated populations following Eqs.~(\ref{eq:fSTIRAP_pops_a}-\ref{eq:fSTIRAP_pops_c}). The inset shows the temporal structure of the Stokes and pump pulses. \label{fig:fstirap}}
\end{figure}

Measured relative populations after f-STIRAP are plotted in Fig.~\ref{fig:fstirap} as a function of ratio $\eta$.
The populations are well described by Eqs.~(\ref{eq:fSTIRAP_pops_a}-\ref{eq:fSTIRAP_pops_c}).
As in the case of STIRAP, 40\,\% of the metastable atoms are lost during the f-STIRAP process due to excitation and spontaneous decay of $^3$D$_2$.

Another interesting system could be realized by using a different polarization scheme of the laser beams, exciting $\sigma^+$- and $\sigma^-$-transitions with Stokes and pump light, respectively.
In this case, STIRAP would lead to a transfer $\ket{+2}\rightarrow\ket{-2}$, and f-STIRAP would allow for the preparation of coherent superpositions in the system ($\ket{+2}$, $\ket{0}$, $\ket{-2}$).
However, this polarization scheme was not experimentally investigated within this work.

\section{Coherence properties \label{sec:coherence}}
The methods presented in Secs.~\ref{sec:5level}-\ref{sec:3level} allow for the preparation of \emph{coherent} superpositions of different $m_J$\hbox{-}states.
In this section, the coherence properties are analyzed using Ramsey~\cite{Ramsey1980} and spin echo~\cite{Hahn1950} experiments.
It is shown that dephasing takes place on a timescale of several 100\,$\mu$s which can be explained by the motion of the atoms along a magnetic field gradient.

As described in Sec.~\ref{sec:5level}, the time evolution of the population of the $m_J$\hbox{-}states can be described by a rotation of the spin.
Particularly, free evolution in a constant magnetic field leads to a rotation about $\vec{e}_z$, and application of a resonant RF field leads to a rotation about $\vec{e}_x$, where $\pi$- and $\pi/2$-pulse are of special importance. (For simplicity, it will not be distinguished between laboratory and rotating frame of reference in the following).

\subsection{Ramsey experiments \label{sec:ramsey}} 
\begin{figure} 
  \includegraphics[width=\linewidth]{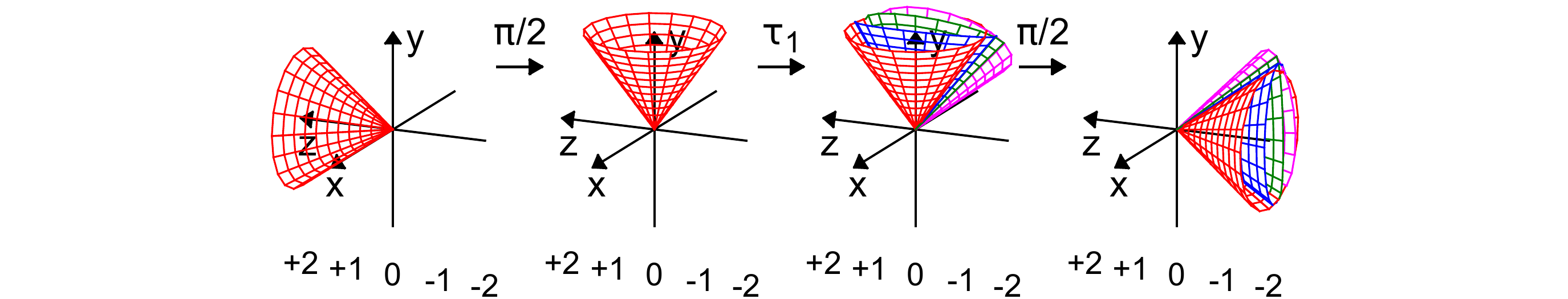}
  \includegraphics[width=\linewidth]{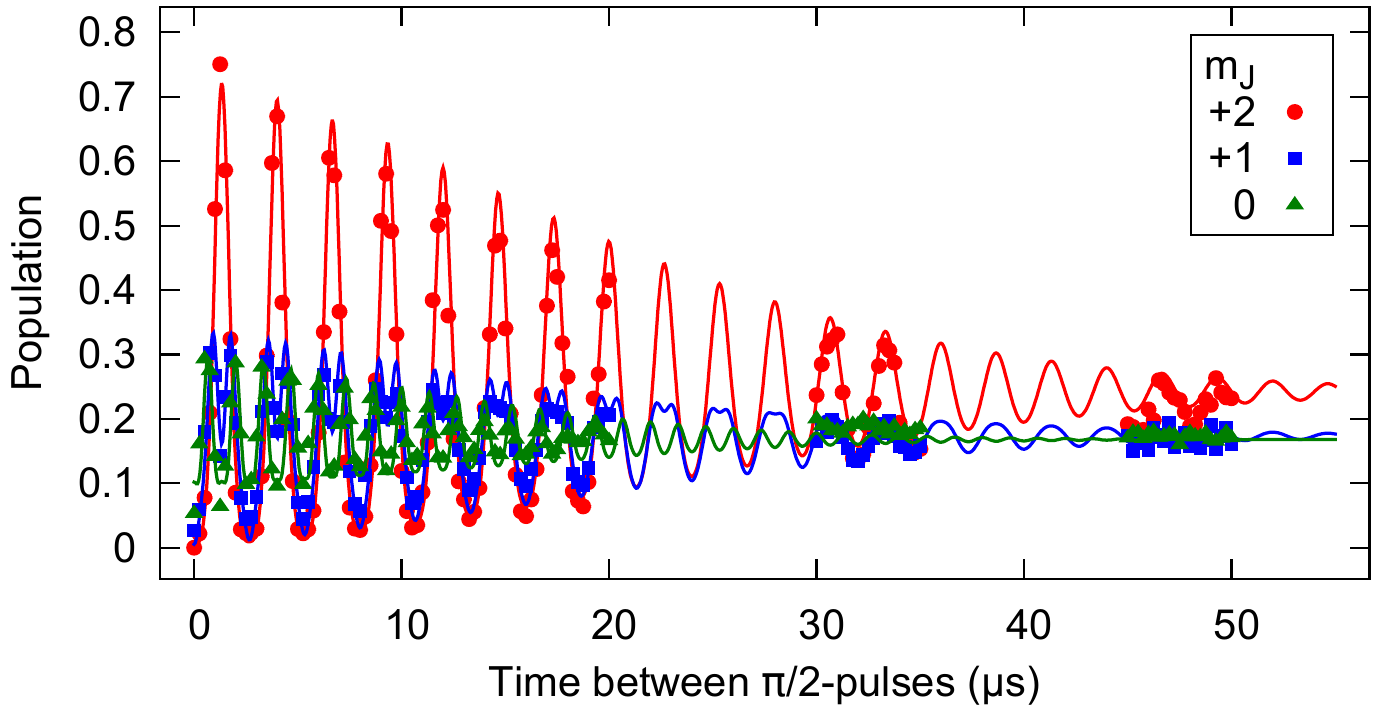}
  \caption{%
(top) Illustration of a Ramsey experiment. The spins are flipped into the x-y-plane by a $\pi/2$-pulse and rotate about the z-axis during the time of free evolution $\tau_1$.
The rotation angle $\Phi$ is measured by application of a second $\pi/2$-pulse, followed by a Stern-Gerlach measurement.
Depending on the local magnetic field, the spins of different atoms rotate by different angles as depicted by the various colored spin cones.
(bottom) Relative populations after a Ramsey experiment as a funcition of the time between the two $\pi/2$-pulses and fitted functions. For simplicity, only populations with $m_J\geq0$ are shown. 
	\label{fig:ramsey}}
\end{figure}

The sequence of a Ramsey experiment~\cite{Ramsey1980} is sketched in Fig.~\ref{fig:ramsey} (top). Initially in state $\ket{\Psi_0}=\ket{+2}$, the spins are flipped into the x-y-plane by a $\pi/2$-pulse. In the following time of free evolution $\tau_1$, the spins rotate by an angle
\begin{equation}
  \Phi(\tau_1)=\int_0^{\tau_1}\dd t\,\gamma\,B\,,
	\quad\tn{with}\quad \gamma=g_J\mu_B/\hbar\,,
\end{equation}
about $\vec{e}_z$.
To determine $\Phi$, a second $\pi/2$-pulse is applied and the projection to $\vec{e}_z$ is measured in a Stern-Gerlach experiment. The resulting relative populations can be calculated using the rotation operators $\op{D}_{\tilde{x},\tilde{z}}$,
\begin{equation}\label{eq:ramsey_pop}
\fl p_{m_J}(\tau_1) = \left|\Bra{m_J} 
\op{D}_{\tilde{x}}\lkk\frac{\pi}{2}\rk
\op{D}_{\tilde{z}}\lkk\Phi(\vphantom{\frac{\pi}{2}}\tau_1)\rk
\op{D}_{\tilde{x}}\lkk\frac{\pi}{2}\rk
\Ket{\Psi_0}\right|^2\hspace{-1ex}.\nonumber
\end{equation}
Particularly, $p_{-2}(\tau_1)=1$ in case of $\Phi(\tau_1)=N\times2\pi$, and $p_{+2}(\tau_1)=1$ in case of $\Phi(\tau_1)=(2N+1)\times\pi$, with $N\in\mathbb{N}$.

In an ensemble of atoms, the rotation angle after evolution time $\tau_1$ will vary from atom to atom, e.\,g. due to a spatial variation of the magnetic field
(see Fig.~\ref{fig:ramsey}, top).
Thus, the amplitude of the oscillation of relative populations, that is measured by variation of $\tau_1$, decreases with increasing $\tau_1$.
For large $\tau_1$, the spin cones fill the whole x-y-plane (x-z-plane after the second $\pi/2$-pulse) which corresponds to the equilibrium populations $p_{+2}=p_{-2}=35/128$, $p_{+1}=p_{-1}=5/32$, and $p_0=9/64$.
If we assume a magnetic field with gradient $B_1$ in z-direction, i.\,e. $\vec{B}=B_0\,\vec{e}_z+B_1\,z\,\vec{e}_z$, the spin of an atom that is initially at position $z_0$ and moves with velocity $v_z$ along $\vec{e}_z$ is rotated by an angle
\begin{eqnarray}\label{eq:ramsey_phi}
  \Phi(\tau_1) 
	&=& \int_0^{\tau_1}\ddd t\lek\gamma\,B_0+\gamma\,B_1\,(z_0+v_z\,t)\rek\\
  &=& \gamma\,B_0\,\tau_1 + \gamma\,B_1\lk z_0\,\tau_1+\frac{1}{2}v_z\,\tau_1^2\rk
\end{eqnarray}
about $\vec{e}_z$ within the free evolution.
For ballistically expanding atomic ensembles with temperature $T$, released from the magnetic trap, the distribution of $z_0$ and $v_z$ is well described by a Gaussian $P(z_0)$, with standard deviation $\sigma_{z,0}$, and a Maxwell-Boltzmann distribution $P_{\tn{MB}}(v_z)$, respectively. The Ramsey signal is given by the ensemble average over Eq.~\eref{eq:ramsey_pop},
\begin{equation}\label{eq:ramsey_average}
  \average{p_{m_J}(\tau_1)} = \int\ddd v_z \int\ddd z_0
	\ p_{m_J}(\tau_1)\,P(z_0)\,P_{\tn{MB}}(v_z)\,.
\end{equation}
The resulting, damped, oscillating terms are of form
\begin{eqnarray}\label{eq:ramsey_dephase}
	\fl\average{\cos\lek\Phi(\tau_1)\rek}\nonumber
	&=& \cos\lk\gamma B_0 \tau_1\rk \,
	\exp\lkk-\frac{1}{2} \gamma^2 B_1^2\sigma^2_{z,0} \tau_1^2\rk\\
  && \times\exp\lkk-\frac{1}{8} \gamma^2 B_1^2\frac{k_B T}{m} \tau_1^4\rk,
\end{eqnarray}
where the first exponential term corresponds to dephasing due to the initial spread of the ensemble and the second due to the motion of the atoms along the gradient.

A measured Ramsey signal is plotted in Fig.~\ref{fig:ramsey} (bottom) as a function of $\tau_1$.
Analytic solutions of Eq.~\eref{eq:ramsey_average} were fitted to the data by variation of $B_1$ and the initial populations $p_{m_J}(0)$.
With a bias field of $B_0=179\,\tn{mG}$ $(\omega_0=2\pi\times375\un{kHz})$, a gradient of $B_1=4.5\un{mG/mm}$ $(2\pi\times9.5\un{kHz/mm})$ is determined.
The oscillation amplitude falls below $1/e$ of the initial value at $\tau_1=32.5\,\mu\tn{s}$.
The fit shows that the signal decrease is dominated by the first exponential function in Eq.~\eref{eq:ramsey_dephase}, i.\,e. by the initial spread of the ensemble with $\sigma_{z,0}=0.73\un{mm}$.
The signal decrease due to the motion of the atoms (i.e. second exponential function in Eq.~\eref{eq:ramsey_dephase}) is significantly smaller on timescales of $\tau_1$ for the given temperature $T_z=0.2\un{mK}$.

\subsection{Spin echo \label{sec:echo}} 
\begin{figure} 
  \includegraphics[width=\linewidth]{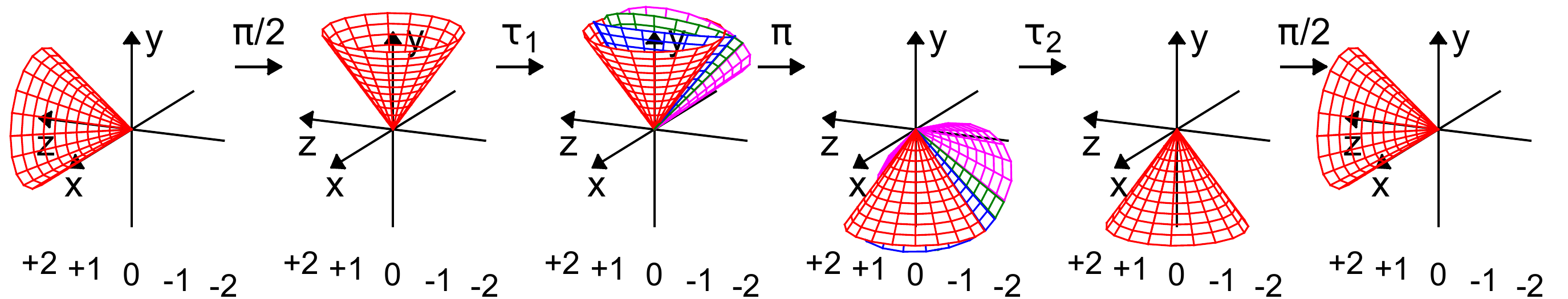}
  \includegraphics[width=\linewidth]{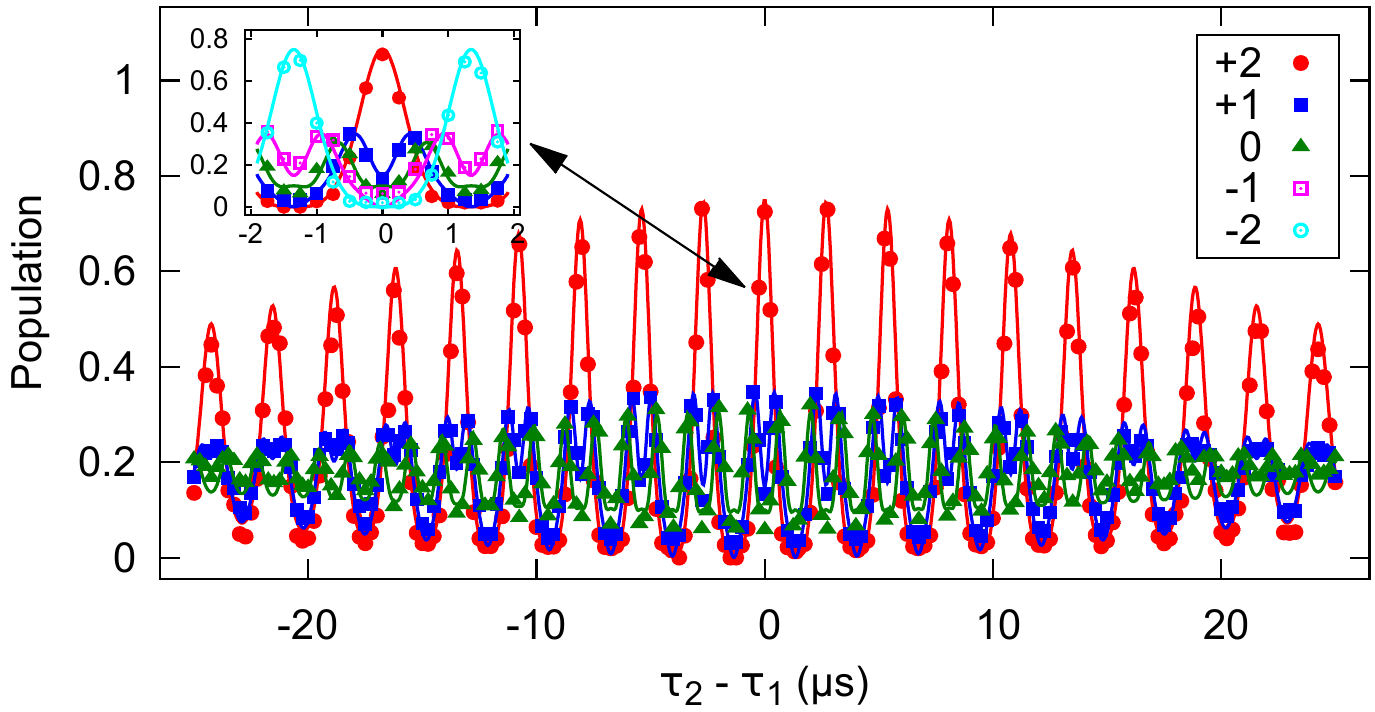}
  \caption{
(top) Illustration of a spin echo experiment. In addition to the Ramsey experiment, a $\pi$-pulse is applied after $\tau_1$ that is followed by a second time of free evolution $\tau_2$ which allows for a rephasing of the spins.
(bottom) Measured relative populations and corresponding fits for a spin echo experiment with $\tau_1=25\,\mu$s and varying $\tau_2$. For simplicity, in the main figure only populations with $m_J\geq0$ are plotted. The inset shows the population for all $m_J$-states close to $\tau_2-\tau_1=0$. At $\tau_2-\tau_1=0$, the oscillation amplitude reaches a maximum value that is called spin echo. 
	\label{fig:echo}}
\end{figure}
The decrease of the Ramsey signal amplitude is due to the differing phase evolution of the different spins in the atomic ensemble. 
For each single atom, however, the phase evolution is deterministic in case of the dephasing due to a magnetic field gradient, as described in the previous section.

The phase evolution can be reverted in a spin echo experiment~\cite{Hahn1950}, the sequence of which is sketched in Fig.~\ref{fig:echo} (top).
As an extension of the Ramsey experiment, an additional $\pi$-pulse is applied after $\tau_1$.
The resulting spin rotation by $\pi$ about $\vec{e}_x$ reverts the order of the spins that have rotated by different angles within $\tau_1$. 
The following additional time of free evolution $\tau_2$ leads to a rephasing of the spins.
As long as the spin evolution is identical within both evolution times, the rotation angles of all spins agree at $\tau_1=\tau_2$ (see Fig.~\ref{fig:echo}, top).

Analogous to Eq.~\eref{eq:ramsey_phi}, the spins rotate by an angle
\begin{eqnarray}
  \fl\Phi(\tau_1, \tau_2)\ &=&\ \int_0^{\tau_1}\ddd t\lek\gamma\,B_0+\gamma\,B_1\,(z_0+v_z\,t)\rek\\
  &&\ -\ \int_{\tau_1}^{\tau_1+\tau_2}\ddd t\lek\gamma\,B_0+\gamma\,B_1\,(z_0+v_z\,t)\rek\nonumber\\
  &=&\ -\gamma\,B_0\,(\tau_2-\tau_1)\ -\ \gamma\,B_1\,z_0(\tau_2-\tau_1)\nonumber\\
  &&\ +\ \frac{1}{2}\gamma\,B_1\,v_z\lek(\tau_2-\tau_1)^2-2\tau_2^2\rek\nonumber
\end{eqnarray}
within $\tau_1$ and $\tau_2$. The resulting relative populations after the spin echo experiment are given by
\begin{eqnarray}\label{eq:echo_states}
  \fl
  p_{m_J}(\tau_1, \tau_2)\hspace{-9ex}&&\\
	&=&
    \left|\Bra{m_J} 
    \op{D}_{\tilde{x}}\lkk\frac{\pi}{2}\rk
    \op{D}_{\tilde{z}}\lkk\vphantom{\frac{\pi}{2}}\Phi(\tau_2)\rk
		\right.\nonumber\\
		&&\hspace{3.5ex}
		\times\left.
		\op{D}_{\tilde{x}}\lkk\vphantom{\frac{\pi}{2}}\pi\rk
    \op{D}_{\tilde{z}}\lkk\vphantom{\frac{\pi}{2}}\Phi(\tau_1)\rk
    \op{D}_{\tilde{x}}\lkk\frac{\pi}{2}\rk
    \Ket{\Psi_0}\right|^2\nonumber\\
    &=&
    \left|\Bra{m_J} 
    \op{D}_{\tilde{x}}\lkk3\,\frac{\pi}{2}\rk
    \op{D}_{\tilde{z}}\lkk\vphantom{\frac{\pi}{2}}\Phi(\tau_1, \tau_2)\rk
    \op{D}_{\tilde{x}}\lkk\frac{\pi}{2}\rk
    \Ket{\Psi_0}\right|^2,\nonumber
\end{eqnarray}
and the ensemble averaged damped oscillating terms are of form
\begin{eqnarray}\label{eq:echo_dephase}
  \average{\cos\lek\Phi(\tau_1, \tau_2)\rek}
	&=&
    \cos\lek\gamma B_0 (\tau_2-\tau_1)\rek\\
		&&\hspace*{-7ex}
    \times\exp\lkk-\frac{1}{2}\gamma^2 B_1^2 \sigma_{z,0}^2(\tau_2-\tau_1)^2\rk\nonumber\\
    &&\hspace*{-7ex}\times\exp\lkk-\frac{1}{8}\gamma^2 B_1^2 \frac{k_B T_z}{m} \lek(\tau_2-\tau_1)^2-2\tau_2^2\rek^2\rk.\nonumber
\end{eqnarray}
Again, the first exponential term corresponds to dephasing due to the initial spread of the ensemble and the second term to dephasing due to the motion of the atoms along the gradient.

As can be seen in Eq.~\eref{eq:echo_dephase}, the difference in the rotation angles that is solely due to the spacial spread of the ensemble (first exponential) is canceled at $\tau_1=\tau_2$.
However, due to their motion, the atoms are exposed to different magnetic fields during the two evolution times.
Thus, the different spin evolution cannot be reversed completely.
Varying $\tau_2$, at constant $\tau_1$, an oscillation of the relative populations is observed, with a maximum amplitude at $\tau_2=\tau_1$. 
The measured relative populations after such a spin echo experiment are plotted in Fig.~\ref{fig:echo}, together with the fitted analytic solution of the ensemble average of Eq.~\eref{eq:echo_states}.

\begin{figure}
	\includegraphics[width=\linewidth]{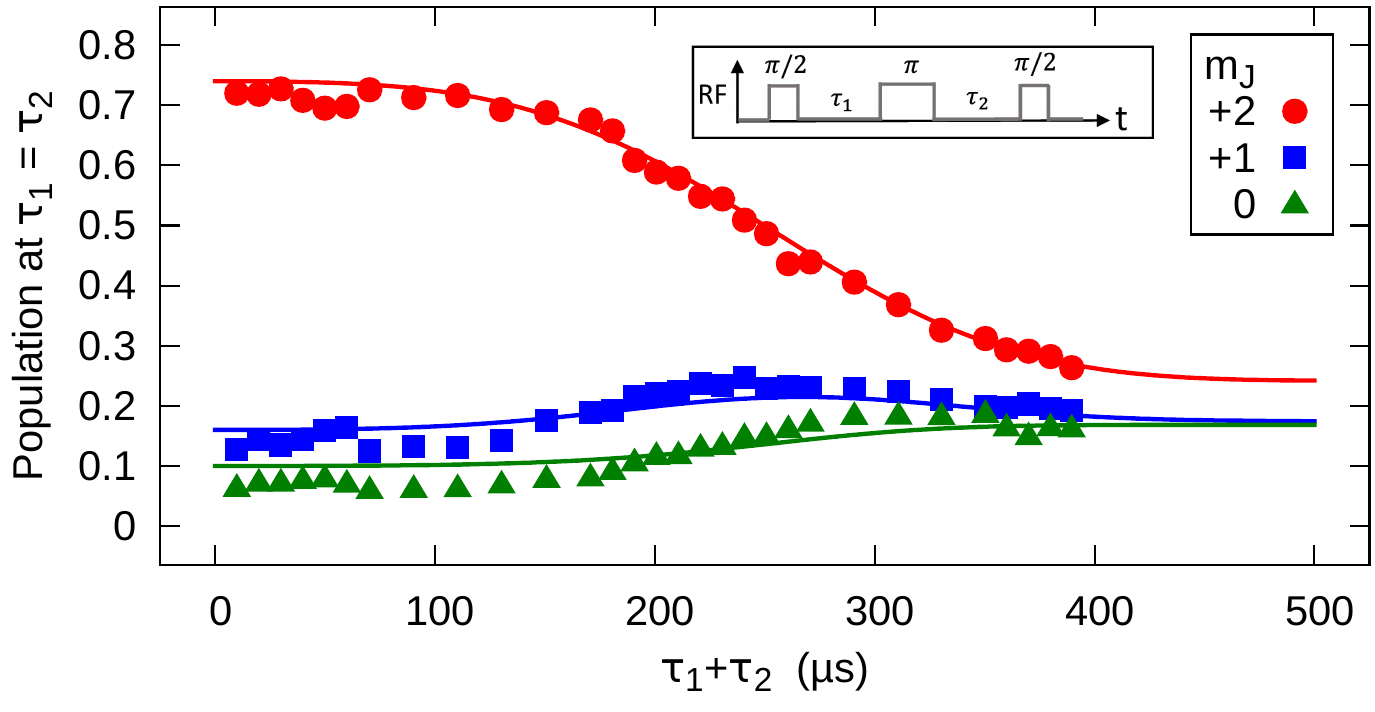}
	\caption{
		Measured relative populations and corresponding fits after spin echo experiments for variation of $2 \tilde{\tau} = \tau_1 +  \tau_2$ with $\tau_1=\tau_2$. For simplicity, only populations with $m_J\geq0$ are shown. The axial temperature $T_z=0.2\un{mK}$ and a magnetic field gradient $B_1=13.5\un{mG/mm}$ $(2\pi\times28.4\un{kHz/mm})$ is determined.
		\label{fig:echo_max}}
\end{figure}

For $\tau_1=\tau_2\equiv\tilde{\tau}$, Eq.~\eref{eq:echo_dephase} becomes
\begin{equation}\label{eq:echo_max_dephase}
  \average{\cos\lek\Phi(\tilde{\tau})\rek} =
  \exp\lk-\frac{1}{2}\gamma^2 B_1^2 \frac{k_B T_z}{m}\tilde{\tau}^4\rk,
\end{equation}
and the signal decrease is of form $e^{-\tilde{\tau}^4}$ for variation of $\tilde{\tau}$.

The results of such a measurement, together with fitted analytic solution of the ensemble average of Eq.~\eref{eq:echo_states}, are plotted in Fig.~\ref{fig:echo_max}.
The signal amplitude (Eq.~\ref{eq:echo_max_dephase}) is reduced to $90\%$ of the initial value after $\tau_1 +  \tau_2=190\un{\mu s}$ and to $50\%$ of the initial value after $\tau_1 +  \tau_2=300\un{\mu s}$.
For large expansion times, the phase correlation between different spins becomes random due to the mixing of the atoms that move with different velocities along the field gradient.
For $\tau_1+\tau_2>500\un{\mu s}$, the ensemble can be considered as an incoherent mixture of spins with relative populations $p_{m_J}$ of states $\ket{m_J}$.
The time that is available to study and apply coherent superpositions, thus, is limited to about 200\un{\mu}s under the given experimental conditions.

Both the Ramsey and the spin echo experiments are very well described by our simple model that assumes a magnetic field gradient along the bias field as the only source of dephasing. This model is consistent with the geometrical arrangements of the bias coils in our experimental setup. 
Due to this good agreement, additional sources of dephasing, such as a field gradient perpendicular to the bias field, temporal drifts of the bias field (typically 5\,mG/100\un{\mu s}), and spatial amplitude variations of the RF field over the ensemble, only have a minor impact on dephasing and have not been considered in greater detail. Decoherence which is considered to be small for superposition states of magnetic sublevels in our experimental configuration of pure magnetic fields in the absence of light can only have a minor impact as well. On the other hand, the results of the Ramsey and spin echo experiments clearly prescribe the path towards an increase of the dephasing time: reduction of the field gradient and of the ensemble temperature and/or confining the atoms to small trapping potentials that are optimized for minimizing additional dephasing or decoherence effects.

\section{Conclusion \label{sec:conclusion}} 
In this work, we performed a detailed investigation of the coherent dynamics of atomic five-level systems. We experimentally demonstrated methods for the robust and efficient preparation of all individual $m_J$-states ($\ket{+2}$, $\ket{+1}$, $\ket{0}$, $\ket{-1}$, $\ket{-2}$) using neon atoms in the metastable $^3$P$_2$ state.
The presented methods can readily be applied to other atomic species, especially those that do not show a hyperfine structure and, thus, lack the adjunct preparation possibilities.

A simple, semiclassical model was introduced to describe the time evolution of the spin states in static magnetic fields and RF fields.
The model is particularly useful for systems with more than two levels and excellently agrees with the measured five-level dynamics.

The presented methods not only allow for the preparation of the individual $m_J$-states but also for the preparation of coherent superposition states in the five-level system ($\ket{+2},\,\ldots,\ket{-2}$), in the two-level system ($\ket{+2},\,\ket{+1}$), and in the three-level system ($\ket{+2},\,\ket{+1},\,\ket{0}$).
The dephasing of the coherent superposition states was analyzed using Ramsey and spin echo experiments and was assigned to the motion of the atoms along a magnetic field gradient. In our setup, the available time for using coherent superpositions is limited to several 100\un{\mu s} which is sufficient for our first application, i.e. the study of effects of the coherent superposition states on the collisional interactions between metastable neon atoms~\cite{Schuetz2013b}.
\ack
We dedicate this article to the late Bruce W. Shore. We thank Bruce for many helpful discussions during all stages of this work. Without his seminal contributions to the field of coherent dynamics in atomic and molecular systems, the results presented in this article would not have been possible. This work has been supported in part by the German Research Foundation (DFG) (Contract No. BI 647/3-1) and by the European Science Foundation (ESF) within the Collaborative Research Project CIGMA of the EUROCORES program EuroQUAM.

\section*{References}
\bibliographystyle{iopart-num}
\bibliography{Schuetz_Superpositions_m-states}

\end{document}